\documentclass[twocolumn]{aastex62}

\usepackage{graphicx}
\usepackage{physics}
\usepackage{booktabs}
\usepackage{nicefrac}
\usepackage{tabularx}
\usepackage{ulem}

\usepackage{savesym}
\savesymbol{tablenum}
\usepackage[separate-uncertainty=true,repeatunits=false]{siunitx}
\restoresymbol{SIX}{tablenum}

\DeclareSIUnit\jansky{Jy}
\DeclareSIUnit\beam{beam}
\DeclareSIUnit\solarmass{M\textsubscript{\(\odot\)}}
\DeclareSIUnit\parsec{pc}
\DeclareSIUnit\yr{yr}
\DeclareSIUnit\gauss{G}
\DeclareSIUnit\erg{erg}

\graphicspath{{./}{figures/}}

\received{27 October 2020}
\revised{21 January 2021}
\accepted{3 February 2021}
\submitjournal{ApJ}

\shorttitle{Toot toot, all aboard!}
\shortauthors{Hodgson et al.}

\begin{document}

\title{Ultra-Steep Spectrum Radio `Jellyfish' Uncovered in Abell 2877}

\correspondingauthor{Torrance Hodgson}
\email{torrance@pravic.xyz}

\author[0000-0003-3443-7123]{Torrance Hodgson}
\affiliation{International Centre for Radio Astronomy Research\\
Curtin University, 1 Turner Ave, Bentley, 6102, WA, Australia}

\author[0000-0001-7703-9040]{Iacopo Bartalucci}
\affiliation{INAF-IASF \\
Via Alfonso Corti 12, 20133 Milano, Italy}

\author[0000-0003-2756-8301]{Melanie Johnston-Hollitt}
\affiliation{International Centre for Radio Astronomy Research\\
Curtin University, 1 Turner Ave, Bentley, 6102, WA, Australia}
\affiliation{Curtin Institute for Computation\\
Curtin University, GPO Box U1987, Perth, 6845, WA, Australia}

\author[0000-0002-9006-1450]{Benjamin McKinley}
\affiliation{International Centre for Radio Astronomy Research\\
Curtin University, 1 Turner Ave, Bentley, 6102, WA, Australia}
\affiliation{ARC Centre of Excellence for All Sky Astrophysics in 3 Dimensions (ASTRO3D), Bentley, Australia}

\author[0000-0002-2821-7928]{Franco Vazza}
\affiliation{Dipartimento di Fisica e Astronomia, Universit\a'a di Bologna, Via Gobetti 92/3, 40121, Bologna, Italy}
\affiliation{Hamburger Sternwarte, Gojenbergsweg 112, 21029 Hamburg, Germany}
\affiliation{INAF, Istituto di Radio Astronomia di Bologna, Via Gobetti 101, 40129 Bologna, Italy}

\author{Denis Wittor}
\affiliation{Hamburger Sternwarte, Gojenbergsweg 112, 21029 Hamburg, Germany}





\begin{abstract}

We report on the discovery of a mysterious ultra-steep spectrum (USS) synchrotron source in the galaxy cluster Abell 2877. We have observed the source with the Murchison Widefield Array at five frequencies across \SIrange[]{72}{231}{\mega \hertz} and have found the source to exhibit strong spectral curvature over this range as well the steepest known spectra of a synchrotron cluster source, with a spectral index across the central three frequency bands of $\alpha = -5.97^{+0.40}_{-0.48}$. Higher frequency radio observations, including a deep observation with the Australia Telescope Compact Array, fail to detect any of the extended diffuse emission. The source is approximately \SI{370}{\kilo \parsec} wide and bears an uncanny resemblance to a jellyfish with two peaks of emission and long tentacles descending south towards the cluster centre. Whilst the `USS Jellyfish' defies easy classification, we here propose that the phenomenon is caused by the reacceleration and compression of \textit{multiple} aged electron populations from historic active galactic nucleus (AGN) activity, so-called `radio phoenix', by an as yet undetected weak cluster-scale mechanism. The USS Jellyfish adds to a growing number of radio phoenix in cool-core clusters with unknown reacceleration mechanisms; as the first example of a polyphoenix, however, this implies the mechanism is on the scale of the cluster itself. Indeed, we show that in simulations, emission akin to the USS Jellyfish can be produced as a short-lived, transient phase in the evolution of multiple interacting AGN remnants when subject to weak external shocks.

\end{abstract}



\section{Introduction}
\label{sec:intro}

Synchrotron sources typically exhibit power law behaviour in their spectra, such that the observed flux $S$ is related to frequency $\nu$ by the relation $S \propto \nu^\alpha$. The spectral index, $\alpha$, is typically around -0.7 for the lobes of active galactic nuclei (AGNs) and other mechanisms, such as radio halos, relics, and AGN remnants, have been observed to have spectra as steep as -2. For most of these sources, the spectral index is a well understood parameter that results from the original injection energy and the aging dynamics of the system.

With the advent of new low frequency telescopes, in particular the Murchison Widefield Array (MWA; \citealp{Tingay2013}) and the LOw Frequency ARay (LOFAR; \citealp{vanHaarlem2013}), it was widely expected that we would uncover a large, hitherto undetected population of ultra-steep spectrum (USS) sources only detectable at low frequency (e.g. \citealp{Ensslin2002,Cassano2012,Cassano2015,vanWeeren2019}). It was believed that traditionally higher frequency radio observations introduced an observational bias against sources whose spectra rapidly declined in luminosity with frequency. But these expectations have not been realised. Instead, it appears that the conditions required to produce low frequency, USS synchrotron emission are not particularly common. And for those that we have found, it is all the more important for us to understand the unique conditions that make them possible.

To date, the steepest reported source is in Abell 1033, the so-called `Gently ReEnergised Tail' (GReET; \citealp{deGasperin2017}) with an integrated spectral index $\alpha =$ \SI{-3.86(3)}{}. The GReET is composed of ancient plasma originally ejected from a wide angle tail radio galaxy, and since reaccelerated. The reacceleration mechanism remains unknown, but the authors considered two scenarios: adiabatic compression due to weak shocks, or stochastic reacceleration driven by complex turbulence in the tail and interaction with the surrounding intracluster medium (ICM). In this particular instance, they concluded that the latter scenario was more likely. The spectral steepness of the GReET was so steep that it was only visible at low frequency and became undetectable above \SI{323}{\mega \hertz}.

The GReET is a kind of radio phoenix, which is a class of synchrotron sources that arises from the reacceleration of ancient but still mildly relativistic ($\gamma > 100$) `fossil' electron populations, usually old AGN cocoons or remnants. Other examples include phoenix in Abell 1664 (e.g. \citealp{Kale2012}), in Abell 2256 \citep{vanWeeren2009}, and recently in Abell 1914 \citep{Mandal2019}. In radio phoenix, the underlying fossil electron population is not well mixed with the ICM and their morphology usually traces out the underlying AGN lobes or tails, albeit made more complex due to diffusion, buoyancy effects, turbulence, and the reacceleration mechanism itself. Radio phoenix rely on a mechanism to reignite an otherwise aged and faded AGN remnant, and adiabatic compression has been suggested as one of these mechanisms \citep{Ensslin2001}, possibly caused by cluster-cluster interaction. In most radio phoenix to date, however, there is no evidence of shocks, and some are even found in relaxed cool-core clusters, implying that major merger events are not required as a trigger. Other mechanisms have been suggested, such as the proposed cool-core `sloshing' in the Ophiuchus cluster to account for its giant radio fossil \citep{Giacintucci2020}. Radio phoenix are typically found close towards the central cluster region and are usually a few hundred kpc in size \citep{Feretti2012}. They often exhibit USS ($\alpha < -1.5$), display spectral curvature, and their spectral index maps typically do not indicate large-scale coherence or trends (see \citealp{vanWeeren2019} and references therein).

Here we report on the discovery of a diffuse USS radio source in the northwest of Abell 2877, also likely a radio phoenix. The source spans $\sim$\SI{740}{\arcsecond} in width, has two bright peaks of emission associated with cluster members, and has tentacles of emission that extend south towards the cluster core, giving the impression of a jellyfish. The `\object{USS Jellyfish}' was first detected in images from the Galactic and Extra-galactic All-sky MWA survey (GLEAM; \citealp{Wayth2015}) by searching for steep spectrum sources. Due to its extreme spectral steepness, it had no detectable counterpart in the highest frequency GLEAM images centred at \SI{200.5}{\mega \hertz} and thus did not form part of the original GLEAM catalogue \citep{HurleyWalker2017}. GLEAM was conducted with the lower resolution `phase I' of the MWA \citep{Tingay2013}, resulting in a blended source that made determining its morphology difficult. This prompted the follow-up radio observations we present here with the higher resolution MWA phase II (extended configuration; \citealp{Wayth2018}) and the Australia Telescope Compact Array (ATCA; \citealp{Frater1992}), as well as reprocessed archival XMM-\textit{Newton} X-ray observations.

Throughout, we assume a flat $\Lambda$CDM cosmology with Hubble constant $h = 0.677$ and matter density $\Omega_m = 0.307$, of which the the baryonic density is $\Omega_b = 0.0486$. All coordinates are with respect to the J2000 epoch. All stated errors indicate one standard deviation.

\subsection{Abell 2877}

Abell 2877, from the southern catalogue of \citet{Abell1989}, is a nearby, low mass cluster in the southern sky at a redshift of $z \approx 0.0238$, an estimated mass $M_{500} =$ \SI{7.103E13}{\solarmass} and radius $r_{500} =$ \SI{0.6249}{\mega \parsec} \citep{MCXC2011}. The cluster was classified by Abell as `poor', having a richness class $R = 0$. It was also earlier catalogued as DC~0107-46 in \citealp{Dressler1980} where it was suggested that it in fact may form a single cluster with nearby Abell 2870 (DC~0103-47) which is centred approximately \SI{4.9}{\mega \parsec} away.

The cluster has two distinct substructures, a core and a substructure to the north, which have been identified from optical data of members of the cluster \citep{Girardi1997, Flin2006}.

Abell 2877 has previously been the subject of radio observations with ATCA at \SI{1.4}{\giga \hertz} as part of the Phoenix Deep Survey \citep{Hopkins2000}. This study detected 15 cluster members at this frequency, of which 14 had spectroscopic observations. These spectra allowed for the classification of six of these galaxies as low-luminosity AGNs, one as a Seyfert 2 galaxy, two as star-forming and the remaining five indeterminately star-forming, low-luminosity AGNs or both. The cluster was otherwise radio-quiet.

\begin{figure*}
    \centering
    \includegraphics[width=\linewidth,clip,trim={0cm 2.5cm 1cm 3.5cm}]{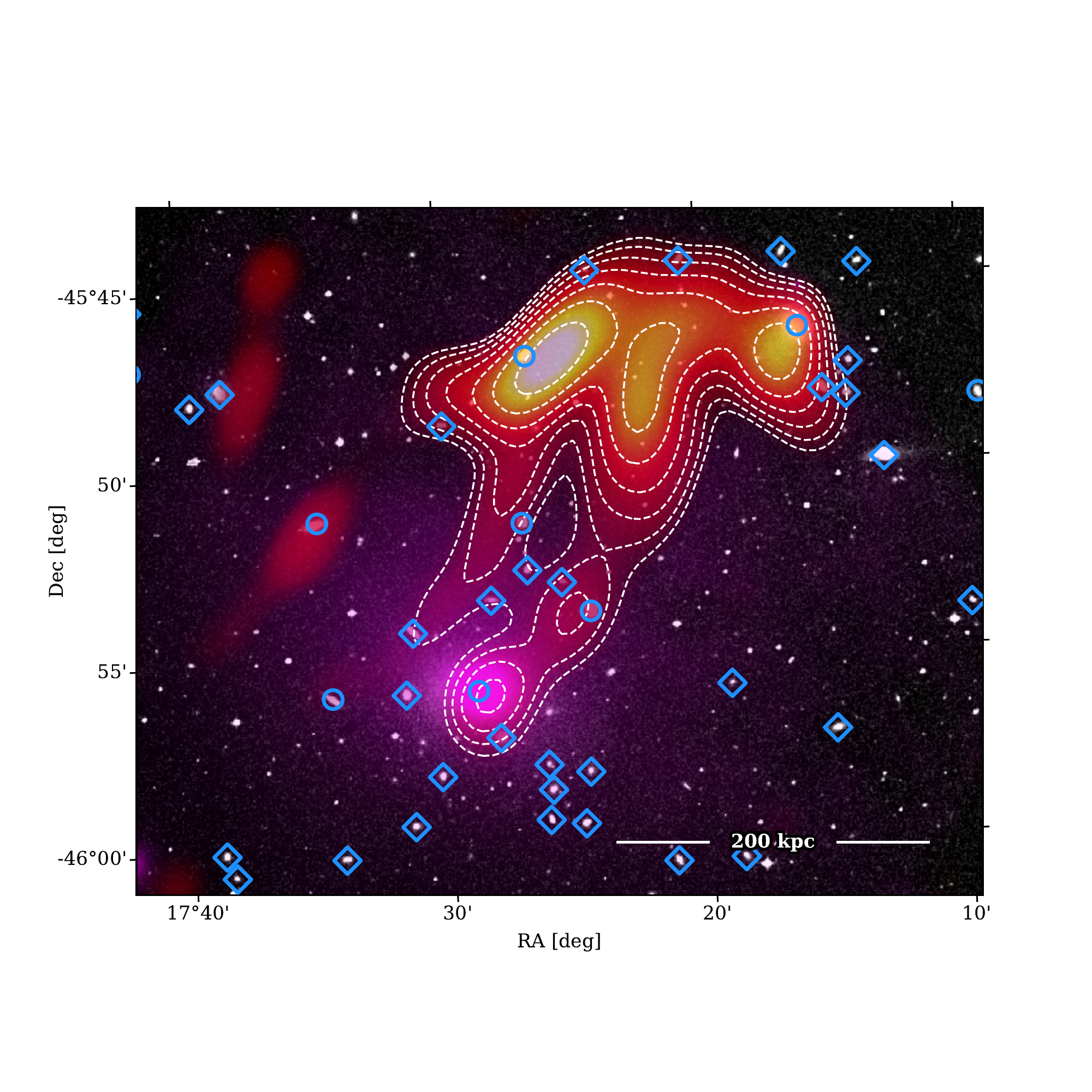}
    \caption{Composite image of the USS Jellyfish in Abell 2877 showing the optical Digitized Sky Survey (background) with XMM X-ray data (magenta overlay) and MWA \SI{118.5}{\mega \hertz} radio data (red-yellow overlay). Contours are provided for the MWA overlay ranging from \SI{7.3}{\milli \jansky \per \beam} and increasing by factors of $\sqrt{2}$, but for clarity, are restricted only to the USS Jellyfish proper. Blue circles indicate cluster radio sources from \citet{Hopkins2000}, and blue diamonds indicate positions of other probable cluster members based on redshift proximity.}
    \label{fig:composite}
\end{figure*}


\section{Data and methods}

\begin{table}[]
    \centering
    \begin{tabular}{p{0.55cm}ccccc} \toprule
        & \textbf{Frequency} & \textbf{Weight} & \textbf{Resolution} & \textbf{PA} & \textbf{Noise} \\ 
        & MHz & & & & \SI{}{\milli \jansky \per \beam} \\ \midrule
        \textbf{MWA} & 87.5 & 0 & \SI{151 x 97}{\arcsecond} & \SI{150}{\degree} & 4.9 \\
        & 118.5 & 0.5 & \SI{123 x 87}{\arcsecond} & \SI{150}{\degree} & 1.9 \\
        & 154.5 & 1 & \SI{108 x 83}{\arcsecond} & \SI{150}{\degree} & 1.3 \\
        & 185.5 & 1 & \SI{93 x 68}{\arcsecond} & \SI{151}{\degree} & 0.91 \\
        & 215.5 & 1 & \SI{77 x 58}{\arcsecond} & \SI{151}{\degree} &  0.86 \\ \midrule
        \textbf{ATCA} & 1548.5 & 0 & \SI{11.8 x 4.1}{\arcsecond} & \SI{15}{\degree} & 0.024 \\
        & 1998.5 & 0 & \SI{10.1 x 3.4}{\arcsecond} & \SI{19}{\degree} & 0.023 \\
        & 2448.5 & 0 & \SI{8.6 x 2.8}{\arcsecond} & \SI{19}{\degree} & 0.027 \\
        & 2899 & 0 & \SI{8.0 x 2.5}{\arcsecond} & \SI{17}{\degree} & 0.030 \\ \bottomrule
    \end{tabular}
    \caption{Radio observations and respective image properties. MWA observations occurred in August 2018 for a duration of 2.4 hours with \SI{30.72}{\mega \hertz} bandwidth. ATCA observations occurred in January 2018 for a duration of 14 hours and have been imaged here with bandwidths of \SI{448}{\mega \hertz}. The baseline weight value refers to the Briggs robustness parameter, and the position angle (PA) of the beam is measured north through east.}
    \label{tab:radioobs}
\end{table}

\subsection{ATCA}

Abell 2877 was observed with ATCA across two observation windows for a total of 11 hours on 16 and 17 January 2018. The observations were undertaken at a central frequency of \SI{2.1}{\giga \hertz} with a bandwidth of just over \SI{2}{\giga \hertz}, however due to radio frequency interference this band was later truncated below \SI{1325}{\mega \hertz}. The observations were conducted in the 750A configuration which consists of baselines ranging from \SIrange[range-phrase=--,range-units=single]{77}{750}{\metre}, with an additional set of baselines ranging in length from \SIrange[range-phrase=--,range-units=single]{3015}{3750}{\metre} produced by the inclusion of the distant, fixed-position sixth antenna. These shortest baselines give angular sensitivity up to scales of $\sim$\SI{9}{\arcminute} at \SI{1.5}{\giga \hertz}, sufficient to detect the large-scale features of the USS Jellyfish.

Initial calibration and flagging were performed with the \texttt{miriad} suite of tools \citep{Miriad1995}. The primary calibrator for each observation window was PKS~1934-638 and a secondary calibrator PKS~0048-427 was observed periodically throughout to ensure phase calibration. We imaged the observation using \texttt{WSClean} \citep{Offringa2014} using the new \texttt{WGridder} backend \citep{Arras2020}. The band was separated into four equally sized output channels using its multifrequency synthesis algorithm \citep{Offringa2017}, baselines were weighted using the Briggs scheme with robustness parameter 0, and Cleaning was performed down to the image noise inside a mask set at a factor of 3 times the noise. Three rounds of phase-only self-calibration were performed using \texttt{Casa} \citep{Casa2007}, before producing the final images (see \autoref{tab:radioobs}).

\subsection{MWA}

\begin{figure*}
    \centering
    \includegraphics[width=\linewidth,clip,trim={2cm 3cm 4cm 3.5cm}]{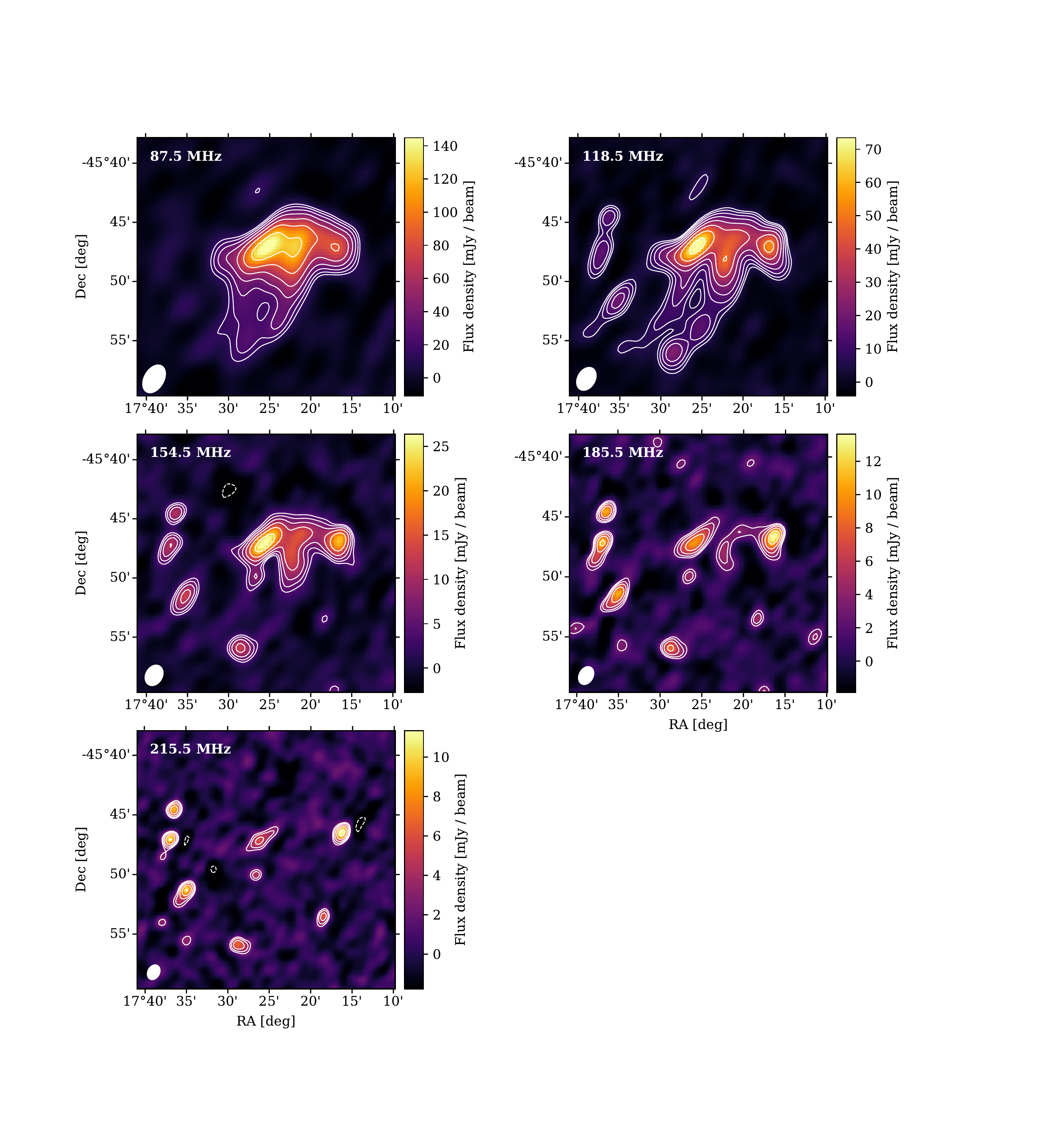}
    \caption{The MWA images of the USS Jellyfish across all five bands. Positive contours start at $3 \sigma$, and increase by factors of $\sqrt{2}$ and are indicated by solid white lines. Negative contours start at $-3 \sigma$, scale similarly, and are indicated by dashed white lines.}
    \label{fig:mwa}
\end{figure*}

The MWA observed Abell 2877 in its phase II extended configuration from 4--14 August 2018. The observations spanned a bandwidth of \SI{30.72}{\mega \hertz} and were centred variously at 87.5, 118.5, 154.5, 185.5, and \SI{215.5}{\mega \hertz}. The cluster was observed in two minute long `snapshots' cumulatively for about 2.4 hours at each frequency band.

All calibration was performed using an in-field model, in contrast to traditional primary calibration workflows. The radio sky model used here was derived from the GLEAM catalogue \citep{HurleyWalker2017} and consisted of the few hundred sources in the field that had an apparent brightness of \SI{1}{\jansky} or more. These sources were predicted into model visibilities based on the primary beam model \citep{Sokolowski2017} and subsequently used as the model to obtain the calibration Jones matrices.\footnote{This work is the first published work to use an updated and significantly faster re-implementation of the MWA calibration software, which is publicly available at \url{https://github.com/torrance/MWAjl}.} Calibration was checked after imaging by source finding on the final mosaics and comparing with the input catalogue; the median flux ratio for each band was \SI{1.00(1)}{} and the spread was $\leq 3$\%.

Imaging was performed using \texttt{WSClean} on individual two-minute snapshots. This is necessitated by the time-dependent phased-array beam of the MWA, such that two minutes is about as long as we can safely assume a constant beam. The images were produced by splitting the \SI{30.72}{\mega \hertz} bandwidth into four equally sized subbands and deconvolved using multiscale Cleaning. Each snapshot was phase rotated to a common phase centre and imaged onto a common projection. The final mosaics were produced by the following process: (i) each of the residual images, the Clean component images, and point spread function (PSF) images were stacked and weighted inversely by the square of the mean noise of the snapshot and the local beam response; (ii) the stacked PSF was fitted with an elliptical Gaussian; and (iii) and this Gaussian was used to convolve the stacked Clean components before being recombined with the stacked residuals.

Imaging and Cleaning on two minute intervals poses significant problems for faint sources. Faint sources remain buried in the noise of individual snapshots and are therefore not Cleaned; they only rise to the level of detection when the residuals are themselves stacked. To mitigate this problem, we performed a final joint deconvolution step on the combined mosaics, performing an image-based Clean with the combined PSF. The effect of this additional Clean procedure was most marked on the lowest band, for which we saw an approximately 20\% increase in recovered flux for the USS Jellyfish; on the higher bands this effect was negligible.

The final weighting, resolution, and noise properties are described in \autoref{tab:radioobs}. The noise properties of these mosaics are the lowest published to date with the MWA and thus provide excellent upper thresholds for detecting the steep radio emission present in Abell 2877.

\subsection{Spectral index and image map}

\begin{table}[]
    \centering
    \begin{tabular}{ccccc} \toprule
        \textbf{RA} & \textbf{Dec} & \textbf{S}$_{\text{\textbf{200}}}$ & \textbf{Spectral index} & \textbf{Label} \\
        \ [deg] & [deg] & [mJy] & \\ \midrule
        17.269 & -45.77 & \SI{12.2(2)}{} & \SI{-0.47(1)}{} & A \\
        17.443 & -45.78 & \SI{5.4(5)}{} & \SI{-0.85(2)}{} & B \\
        17.480 & -45.93 & \SI{6.4(11)}{} & \SI{-0.66(4)}{} & cD \\ 
        17.444 & -45.83 & \SI{4.4(11)}{} & \SI{-0.40(6)}{} & E \\ \bottomrule
    \end{tabular}
    \caption{Radio point sources that overlap with the integrated flux mask. Values are derived from fitting a power law to RACS, ATCA and the \SI{215.5}{\mega \hertz} MWA image.}
    \label{tab:pointsources}
\end{table}

The integrated flux of the USS Jellyfish was calculated by using a mask that extended out to the $1 \sigma$ boundary of the USS Jellyfish in the \SI{87.5}{\mega \hertz} image, and summing the flux for each respective MWA image over this mask. Embedded point sources were subtracted from the integrated sum, where their respective flux density was estimated based on fitting simple power law functions to measurements from all four ATCA bands, the \SI{887}{\mega \hertz} Rapid ASKAP\footnote{Australian Square Kilometre Array Pathfinder.} Continuum Survey mosaic (RACS; \citealp{McConnell2020}), and the \SI{215.5}{\mega \hertz} MWA image (see \autoref{tab:pointsources}). To be included, point sources required a detectable counterpart at the $1.5\sigma$ level in the highest frequency MWA image to help constrain the measurement and to filter sources that exhibited a turnover at higher frequencies. Errors were calculated in quadrature based on: 5\% absolute flux error; 10\% point source error; and local map noise. This latter local map noise dominates the measurements at the highest frequency due to the large area over which the flux is summed and the proportionally faint emission present.

Spectral index values, both multiband and pairwise, have been calculated using a bootstrapping technique with 100,000 samples to propagate errors. Stated values represent the median value over all samples, and errors indicate the 15.9th and 84.1th percentile values. In the case where the lower error bounds produce negative flux values, we have included only the 84.1th percentile value as an upper bound.

The spectral image map was produced from the lowest three MWA bands at 87.5, 118.5 and \SI{154.5}{\mega \hertz}; the higher band images were not used as they contained little of the extended diffuse emission. The snapshots at these three bands were imaged with identical pixel resolution, projection and phase centre so as to avoid having to resample and interpolate later. The images were convolved to the common resolution of the \SI{87.5}{\mega \hertz} image. We then employed a bootstrapping technique with 1,000 samples to estimate the spectral index and associated error for each pixel across the three images. Point source subtraction was not performed.

\subsection{XMM \textit{Newton}}

The XMM-\textit{Newton} observation of Abell 2877 consists of six pointings taken using the European Photon Imaging Camera (\citealt{turner2001,struder2001}), for a total exposure time of \SI{475}{\kilo \second}.\footnote{Observation IDs 0801310101, 0693060401, 0693060301, 0655510201, 0560180901, and 0204540201.} Each observation was reduced by using the Science Analysis System version 18.0.0 and applying the latest calibration files available as of December 2019. We removed from the analysis observation intervals affected by flares following the procedures described in \citealt{pratt2007}, yielding \SI{420}{\kilo \second} of useful time. The pointings were combined to maximise the signal-to-noise ratio. Exposure maps, models of the sky and instrumental background were computed as described in \citet{bourdin08}, \citet{bourdin13} and \citet{bogdan13}. Finally, we identified and removed point sources using the wavelet detection technique of \citealt{bogdan13}.

Unfortunately, the northwest sector of the cluster that hosts the USS Jellyfish is covered by only a single pointing with low exposure time. For this reason, the brightness and temperature maps are not sufficiently deep to investigate the presence of faint or subtle features.

\section{Results}

    
    

\begin{table}[]
    \begin{tabularx}{\columnwidth}{Xcc} \toprule
        \textbf{Frequency} & \textbf{Flux Density} & \textbf{Spectral Index} \\
        MHz & Jy & \\ \midrule
        87.5 & \SI{1.10(06)}{} & \smash{\raisebox{-.5\normalbaselineskip}{$\alpha^{118.5}_{87.5} = -2.16^{+0.27}_{-0.27}$}} \\
        118.5 & \SI{0.57(3)}{} & \smash{\raisebox{-.5\normalbaselineskip}{$\alpha^{154.5}_{118.5} = -4.87^{+0.38}_{-0.39}$}}\\
        154.5 & \SI{0.16(1)}{} & \smash{\raisebox{-.5\normalbaselineskip}{$\alpha^{185.5}_{154.5} = -7.8^{+1.1}_{-1.3}$}}\\
        185.5 & \SI{0.038(8)}{} & \smash{\raisebox{-.5\normalbaselineskip}{$\alpha^{215.5}_{185.5} < -9.2$}}\\
        215.5 & \SI{0.003(7)}{} & \\ \bottomrule
    \end{tabularx}
    \caption{The integrated flux of the USS Jellyfish at each MWA band, as well as pairwise spectral indices. We provide only an upper bound for the highest frequency pairwise spectral index as the integrated flux measurement at \SI{215.5}{\mega \hertz} is consistent with zero.}
    \label{tab:spectralindex}
\end{table}

\begin{figure}
    \centering
    \includegraphics[width=\linewidth,clip,trim={0 0cm 0 -0.5cm}]{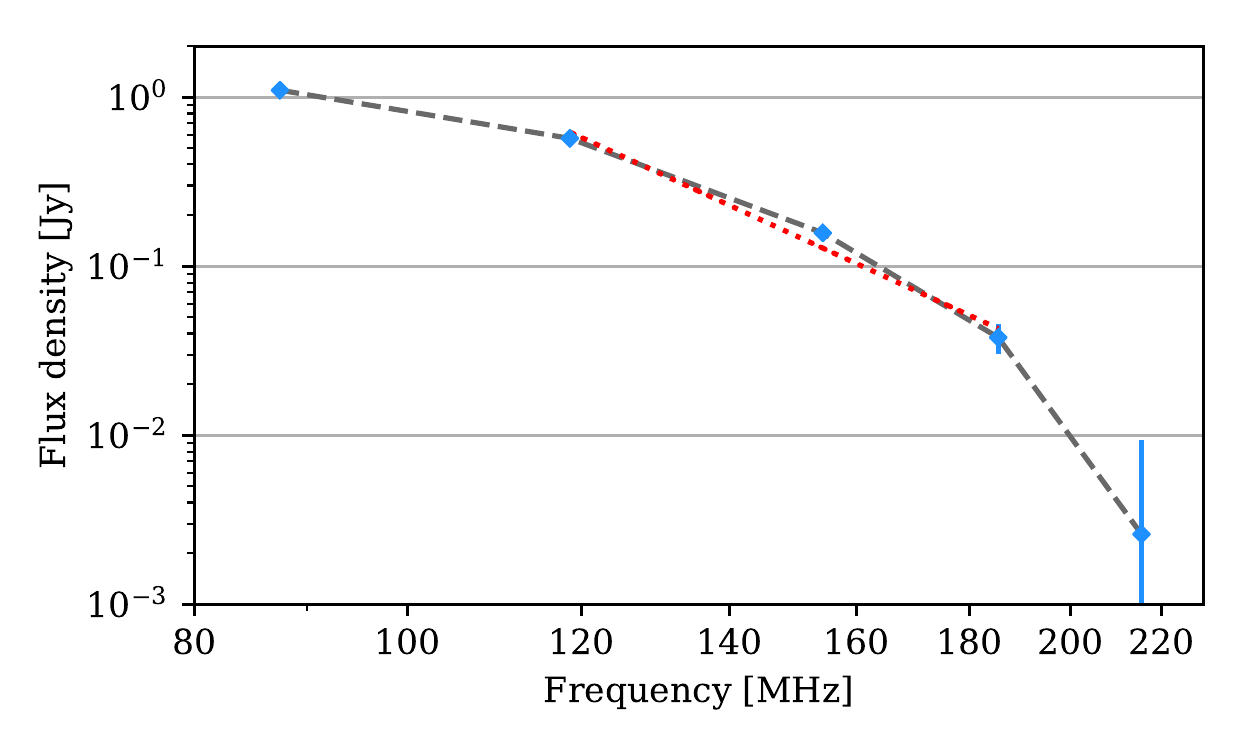}
    \caption{Plotted values of the integrated flux of the USS Jellyfish, which make the curvature apparent. By fitting a power law across the central three frequencies (red), we find a spectral index value of $\alpha = -5.97^{+0.40}_{-0.48}$.}
    \label{fig:spectralindex}
\end{figure}

\begin{figure}
    \centering
    \includegraphics[width=\linewidth,clip,trim={1cm 0 2cm 1.1cm}]{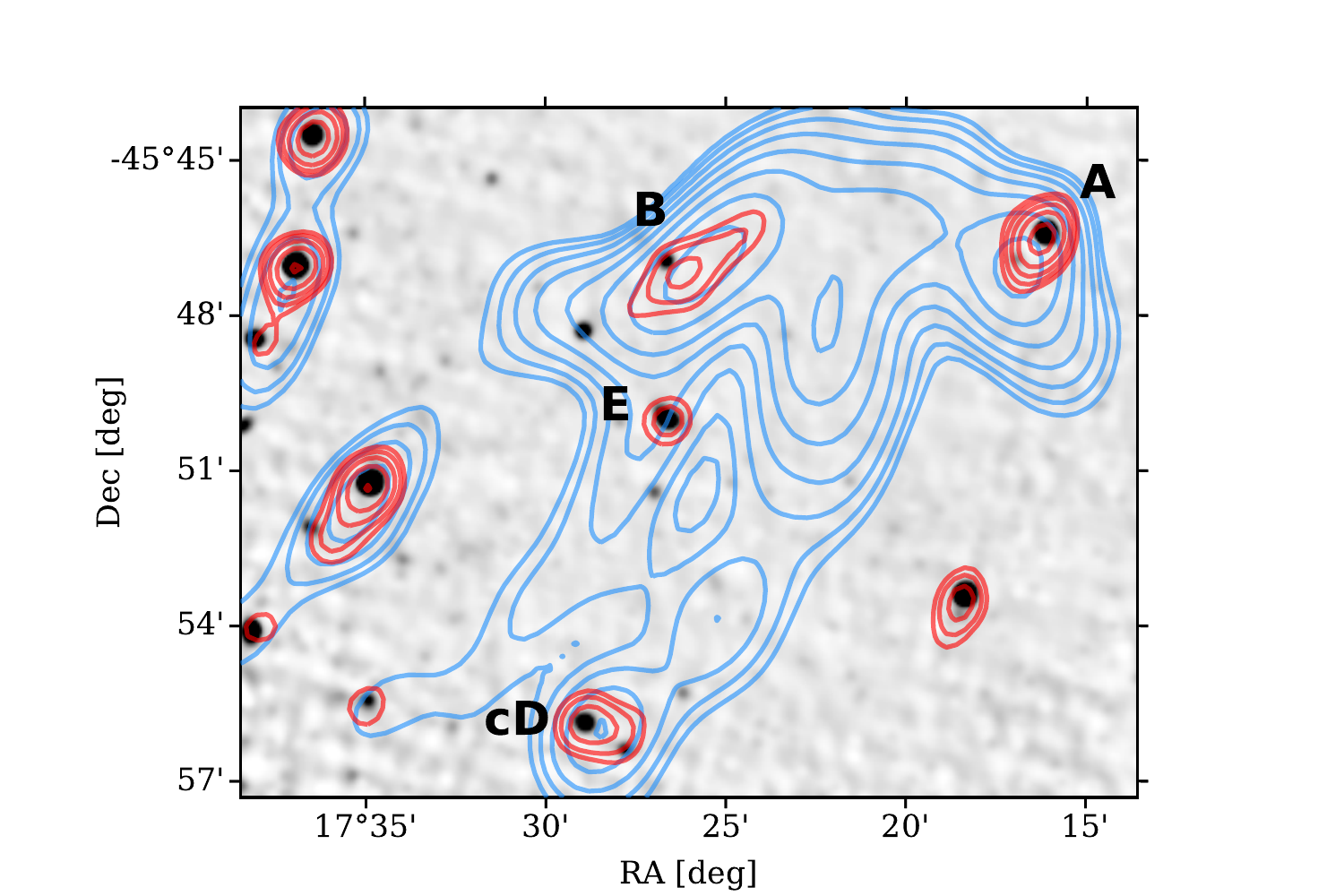}
    \caption{Combined ATCA observation centred at \SI{2223.5}{\mega \hertz}, with full \SI{1.8}{\giga \hertz} bandwidth. For display purposes, it is convolved down to a circular \SI{15}{\arcsecond} resolution, giving a local RMS noise of \SI{28}{\micro \jansky \per \beam}. The colour scale ranges from \SIrange[range-phrase=\ to\ ,range-units=single]{-0.1}{0.5}{\milli \jansky} and is set to saturate the majority of the ATCA sources. Contours show MWA \SI{118.5}{\mega \hertz} (blue) and \SI{215.5}{\mega \hertz} (red). The labels indicate three cluster members: (A) ESO~243~G~045; (B) WISEA~J010946.55-454657.4; (cD) IC~1633. Additionally, (E) indicates the position of a background radio galaxy.}
    \label{fig:atca}
\end{figure}

\begin{figure*}
    \centering
    \includegraphics[width=\linewidth,clip,trim={2.7cm 0 2.5cm 0}]{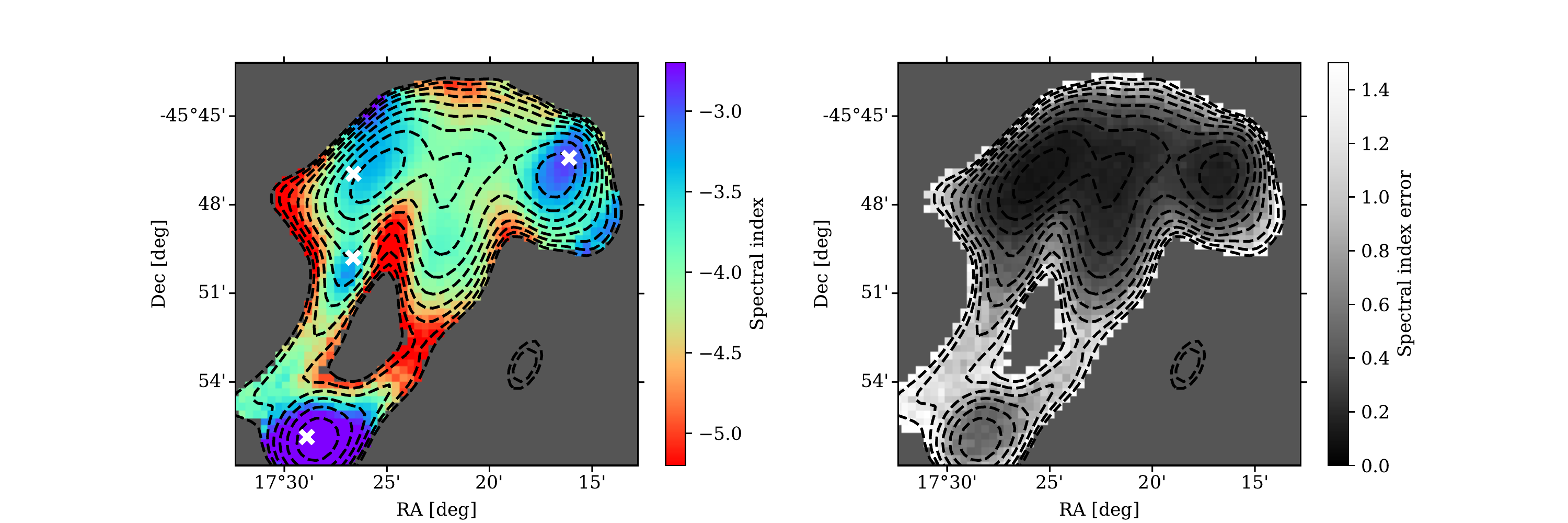}
    \caption{Spectral image map calculated across the lowest three MWA bands, with the locations of point sources from \autoref{tab:pointsources} indicated with white crosses. The map shows cocoons of shallower spectrum emission slightly offset from points A and B, as well as aligned with points cD and E, whilst elsewhere we observe much steeper emission. \textit{Left:} The spectral index values. \textit{Right:} The respective error map generated using a bootstrapping method.}
    \label{fig:spectralmap}
\end{figure*}

We present the USS Jellyfish in \autoref{fig:composite} as a composite image combining the optical Digitized Sky Survey (background) with wavelet cleaned XMM X-ray data (magenta overlay) and MWA \SI{118.5}{\mega \hertz} radio data (red-yellow overlay). Blue circles indicate known radio emitting cluster sources from \citet{Hopkins2000}, whilst blue diamonds indicate probable cluster members based on available redshift data. The X-ray emission shows the core overdensity of Abell 2877, to which the radio emission is offset to the northwest. The radio emission resembles a jellyfish, with western and eastern peaks of emission in the head, and likewise western and eastern tentacles that descend south towards the cluster core. The full angular extent of the USS Jellyfish, from east to west, is $\sim$\SI{740}{\arcsecond} which at the redshift of Abell 2877 corresponds to $\sim$\SI{370}{\kilo \parsec} in projection.

\autoref{fig:mwa} shows the emergence of the USS Jellyfish from near undetectability at \SI{215.5}{\mega \hertz} to an integrated flux of \SI{1.10(15)}{\jansky} at \SI{87.5}{\mega \hertz}, approximately 275 times more luminous. At this lowest frequency, the integrated flux corresponds to a total radio luminosity of $L_{87.5 \text{MHz}} =$ \SI{1.59(22)E24}{\watt \per \hertz}, assuming a redshift $z = 0.0238$. 

The steepness of the spectra is extreme and shows significant curvature, as can be seen in \autoref{fig:spectralindex}. This figure shows the integrated flux measured in each band, linearly interpolated (grey dashed line), whilst \autoref{tab:spectralindex} additionally provides pairwise spectral indices. If we attempt a power law fit ($S \propto \nu^\alpha$) to the three central bands, we find a spectral index value of $\alpha =-5.97^{+0.40}_{-0.48}$.

Unsurprisingly, higher frequency observations show no extended emission that is coincident with the USS Jellyfish. The previously stated spectral index would give a total integrated flux at \SI{887}{\mega \hertz} of just \SI{4}{\micro \jansky}. Thus, at \SI{887}{\mega \hertz}, the respective RACS mosaic, with a local noise of \SI{220}{\micro \jansky \per \beam}, shows no emission besides point source emission from Abell 2877 galaxy members; additional convolution steps to emphasise extended diffuse emission also do not reveal any signal. Similarly, the lowest frequency band of our ATCA observation at \SI{1548.5}{\mega \hertz}, with a local noise of \SI{52}{\micro \jansky \per \beam}, does not reveal any of the extended emission of the Jellyfish, nor do the higher frequency ATCA bands.

There are three radio-emitting members of Abell 2877 entangled in the USS Jellyfish that are detectable in the \SI{215.5}{\mega \hertz} MWA image, and there is a fourth background source. In \autoref{fig:atca} we show the combined ATCA observation centred at \SI{2223.5}{\mega \hertz} overlaid with contours from the MWA at both \SI{215.5}{\mega \hertz} and \SI{118.5}{\mega \hertz}. In the highest frequency MWA image, we identify two sources of emission, labelled A and B, that at lower frequencies are closely associated with the bright western and eastern peaks at the head of the USS Jellyfish. At \SI{215.5}{\mega \hertz}, the radio emission at A and the E/S0 galaxy ESO~243~G~045 are well aligned. There is evidence of a slight southeast elongation of the emission, and indeed, at lower frequencies the western peak shifts southeast slightly from A. Source B is offset slightly to the southwest of S0 galaxy WISEA J010946.55-454657.4. The northwest alignment of the emission of source B at \SI{215.5}{\mega \hertz} remains visible at lower frequencies suggesting continuity of the emission source. \citet{Hopkins2000} had previously classified both of these radio sources: ESO~243~G~045 as a low-luminosity AGN and WISEA J010946.55-454657.4 unambiguously as a Seyfert 2 galaxy. Source cD is the central cluster galaxy, IC 1633, and in the lower frequency bands the western tentacle bridges this source to the rest of the emission. It is identified by \citet{Hopkins2000} as hosting an AGN, and both their ATCA observation and ours show nearby faint emission to the northeast that may indicate a small jet. A background radio galaxy ($z = 0.545$; \citealp{Afonso2005}) with resolved $\sim$\SI{15}{\arcsecond} jets is also visible in the MWA images and is labelled E.

The spectral index map (\autoref{fig:spectralmap}) shows no overall trend in the spectral index values across the full extent of the emission. Instead, we observe a patchwork of islands of shallower emission, surrounded by more diffuse and steeper emission. Both the western and eastern peaks of emission at the head of the USS Jellyfish are associated with flatter spectra, as is the emission coincident with the cD galaxy itself and the background radio galaxy E. Note that the lower spectral index values at the western and eastern peaks are not simply caused by a bias introduced from point sources A and B, since both are offset to their respective galaxy counterparts. The rest of the extended emission has a spectral index of around -4, and at some edges and along the western tentacle, the spectrum tends steeper still.

\begin{figure}
    \centering
    \includegraphics[width=\linewidth,clip,trim={0.4cm 0 0.4cm 0}]{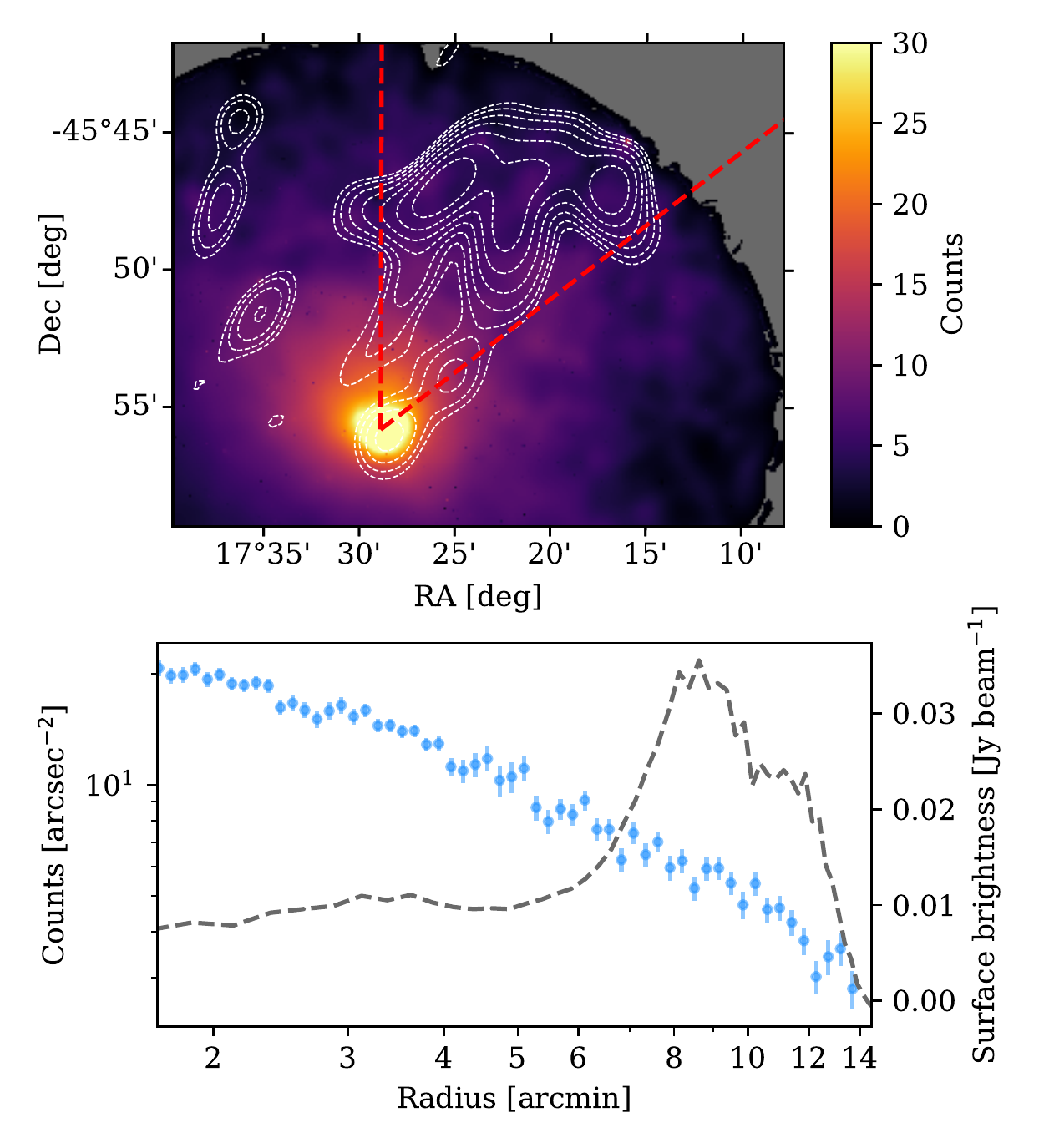}
    \caption{\textit{Top:} Wavelet de-noised map of the XMM-\textit{Newton} image in the \SIrange[range-phrase=--,range-units=single]{0.5}{2.5}{\kilo \electronvolt} band produced following \citealt{bourdin04}, with overlay from MWA \SI{118.5}{\mega \hertz} (dashed white contours). Dashed red lines indicate the extent of the northwest sector. \textit{Bottom:} The X-ray count density (blue points) and the MWA \SI{118.5}{\mega \hertz} mean surface brightness (dashed grey line) throughout the northwest sector as a function of radial distance from the cluster core. There is no indication of a shock or cold front.}
    \label{fig:xmm}
\end{figure}

From the X-ray, we observe Abell 2877 to strongly resemble a cool-core cluster, with the core being X-ray bright and having a temperature of $\sim$\SI{2}{\kilo \electronvolt}. In \autoref{fig:xmm} we show the exposure-corrected and background-subtracted X-ray surface brightness profile of the northwest sector in the \SIrange[range-phrase=--,range-units=single]{0.5}{2.5}{\kilo \electronvolt} band (blue points) alongside the radio surface brightness profile of the USS Jellyfish at \SI{118.5}{\mega \hertz} (dashed grey line). There is no evidence for a shock or cold front in this sector; however, it remains possible that this is due to the poor data in this sector or that the feature is intrinsically faint due to projection effects.

\section{Discussion}

USS emission on cluster-sized scales points to only a handful of plausible mechanisms to account for its spectral steepness: radio relics and halos, AGN remnants, and reaccelerated AGN plasma or `phoenix'.\footnote{See \citet{vanWeeren2019} for an excellent review of each of these phenomena.} We can readily discount halos as the emission is significantly offset from the cluster core, and radio halos trace the baryonic content of the cluster. Radio relics trace large-scale cluster shocks and usually appear towards the periphery of a cluster, although they have been observed more centrally. Relics have values for $\alpha$ of typically -1 to -1.5, and often display large-scale coherency in their spectral index map, indicating the direction of the shock, which we do not observe here. Their morphology is usually long and narrow, as they trace the bow of the shock. The spectral index of a relic is related to the shock strength and, assuming diffusive shock acceleration (DSA) on a purely thermal plasma and a conservatively shallow value of $\alpha_{\text{inj}} < -3$, we find a shock Mach number $\mathcal{M} < 1.3$.\footnote{It should be noted that Mach numbers derived from radio and X-ray typically disagree, and the values derived from radio are overestimated when compared to those derived from X-ray. See review discussion in \citet{vanWeeren2019}.} Modelling of DSA by \citet{Hoeft2007} shows that the efficiency of electron acceleration is strongly dependent on the Mach number, and for shocks $\mathcal{M} < 3$ their models show a rapid and exponential decrease in this efficiency. In the case of the USS Jellyfish, and without recourse to a significant fossil electron population (e.g. see models by \citealp{Kang2011,Pinzke2013}), such a truly gentle shock is an exceptionally inefficient electron accelerator. For these reasons, we deem the radio relic hypothesis unlikely.

An AGN remnant is another potential scenario, as models for synchrotron aging predict a steepening spectral index, increasing spectral curvature and a decreasing `break frequency' (e.g. \citealp{Kardashev1962,Pacholczyk1970} `KP model'; \citealp{Jaffe1973} `JP model'). None have been observed to be as steep as observed here; however, it is possible that we are observing the remnant in a frequency range \textit{above} the break frequency. If we assume the break frequency to be $<$ \SI{70}{\mega \hertz} and the cluster magnetic field strength to be in the range of $\approx$ \SIrange[range-phrase=--,range-units=single]{0.01}{1}{\micro \gauss},\footnote{For example, measurements of the magnetic field in Abell 3667 find values approximately \SIrange[range-phrase=--,range-units=single]{1}{3}{\micro \gauss} \citep{JohnstonHollitt2004,Riseley2015}.The \SI{0.01}{\micro \gauss} lower bound is a conservative value derived from equipartition assumptions of the electric and magnetic field.} we find a synchrotron age for the USS Jellyfish of at least \SIrange[range-phrase=--,range-units=single]{24}{220}{\mega \yr} or \SIrange[range-phrase=--,range-units=single]{53}{490}{\mega \yr}, based on the KP and JP models, respectively. This is significantly older than any known AGN remnant, and indeed, recent simulations suggest rapid cooling and dimming of radio lobes after AGN shutoff due to adiabatic losses on the order of just millions of years \citep{Godfrey2017,English2019}. It is unlikely that the USS Jellyfish is an undisturbed AGN remnant.

Instead, our principal hypothesis is that it is far more likely that the USS Jellyfish is composed of \textit{multiple} radio phoenix---a polyphoenix---triggered by a common large-scale shock or compression occurring in the northwest of Abell 2877. The strongest evidence for this are the western and eastern peaks of emission and their association with nearby point sources A and B, respectively. Both peaks are the least steep components of the spectral image map, indicating a cocoon of younger, more energetic electrons that envelop cluster members ESO~243~G~045 and WISEA J010946.55-454657.4. We suggest that the shallower spectrum component at the western peak, though it is offset to the southeast of source A, is directly related to emission from ESO~243~G~045. We also suggest that the elongated eastern peak is the product of a pair of weak AGN jets originating from source B, WISEA J010946.55-454657.4; the slight offset likely indicates that this activity was historic. Moreover, the eastern spur of the jellyfish head can be explained by deflection of the south eastern jet of source B by the denser, more central ICM. While both electron populations would have rapidly dimmed if unperturbed, we suggest that they have been compressed and reignited by some external mechanism(s).

The large-scale, more diffuse emission of the USS Jellyfish, as well as its tentacles, remains difficult to explain. The eastern extension from ESO~243~G~045 that is visible at \SI{185.5}{\mega \hertz} is suggestive of either a lobe from a previously active epoch of AGN activity or a tail. If it is the latter, one plausible explanation is that both tentacles are the re-energised tail of ESO~243~G~045 that sweeps south towards the core of the cluster. An alternative explanation would be to invoke the cD galaxy IC 1633 as a third electron source, since the western tentacle, and to some degree the eastern, establishes a bridge of emission to this source.

The fact that multiple disparate electron populations have been reaccelerated strongly suggests a common large-scale reacceleration mechanism. As the XMM observation shows, there is presently no evidence of a shock in the northwest sector; moreover, it appears Abell 2877 is a relaxed, cool-core cluster and therefore unlikely currently subject to disruptive merger events or other large-scale structure formation processes that could trigger such a large-scale shock. Cool-core sloshing, which is identified in the X-ray by the presence of spiral or arc-shaped cold fronts about the central core, is an alternative mechanism that is triggered by instabilities acting upon the deep gravitational well at the centre of cool-core clusters \citep[see e.g.,][]{Ascasibar2006,Ghizzardi2010,Vazza2012}. Indeed, the previously identified northern substructure in Abell 2877 may provide just such an instability. Whilst this mechanism has typically been invoked to explain centrally located `mini-halos' \citep[e.g.,][]{Giacintucci2014}, cold fronts associated with cool-core sloshing have been observed up to \SI{1}{\mega \parsec} away \citep{Simionescu2012, Rossetti2013}. However, until we can obtain both high-resolution and high-sensitivity X-ray observations of the cluster, the existence and nature of any shocks in the system must remain purely speculative.

Of special note, the spectral index map of the USS Jellyfish indicates spectra that are still strongly correlated with the original plasma age. In a standard DSA scenario, a strong shock would have imprinted the strength of the shock itself on the spectrum of the plasma and would display the same kind of large-scale coherence observed in the spectral index maps of many relics. We can thus infer a particularly gentle shock, and, based on our current understanding of DSA physics and the inefficiency of weak shocks, we can also infer the existence of a significant population of suprathermal electrons throughout the northwest sector of Abell 2877, further reinforcing the phoenix hypothesis. Additionally, adiabatic compression of AGN cocoons akin to that originally described in \citet{Ensslin2001} is another complimentary mechanism. Indeed, due to the higher speed of sound in the cocoon environment, shock waves will only poorly penetrate the AGN cocoons, and DSA would thus be of negligible effect. Their modelling suggests that adiabatic compression can boost the luminosity of the AGN cocoons without a significant flattening of the original spectrum, thus preserving the underlying aged spectra. Such compression, however, is unlikely to explain the large-scale diffuse radio emission observed exterior to the AGN cocoons. We also raise the possibility of a third reacceleration mechanism, local turbulence in the wake of a weak shock between interacting AGN lobes, powering Fermi II acceleration processes.

\begin{figure}
    \centering
    \includegraphics[width=1\linewidth,clip,trim={1.1cm 0.8cm 0.8cm 1cm}]{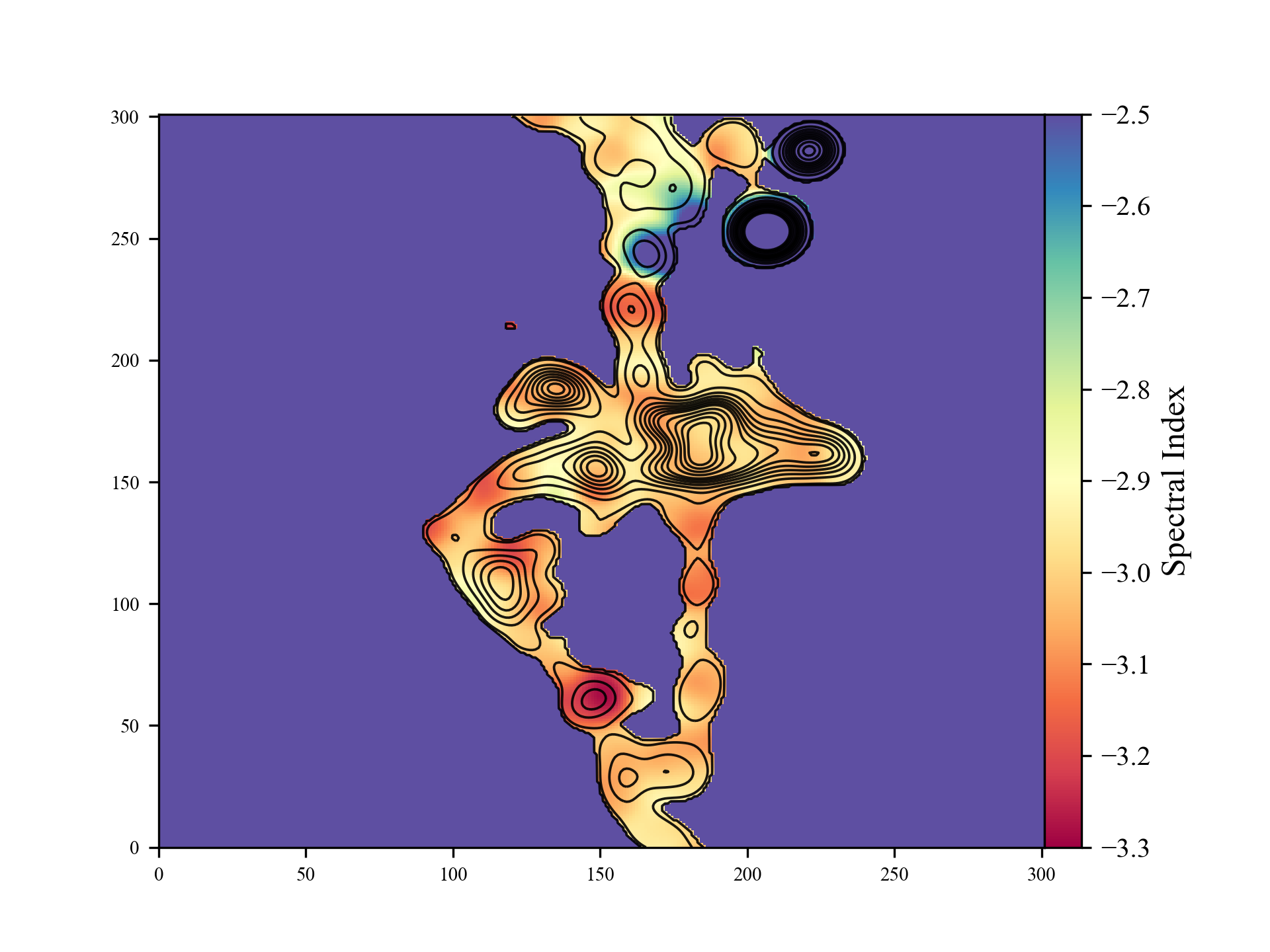}
    \includegraphics[width=1\linewidth,clip,trim={1.1cm 0.8cm 0.8cm 1cm}]{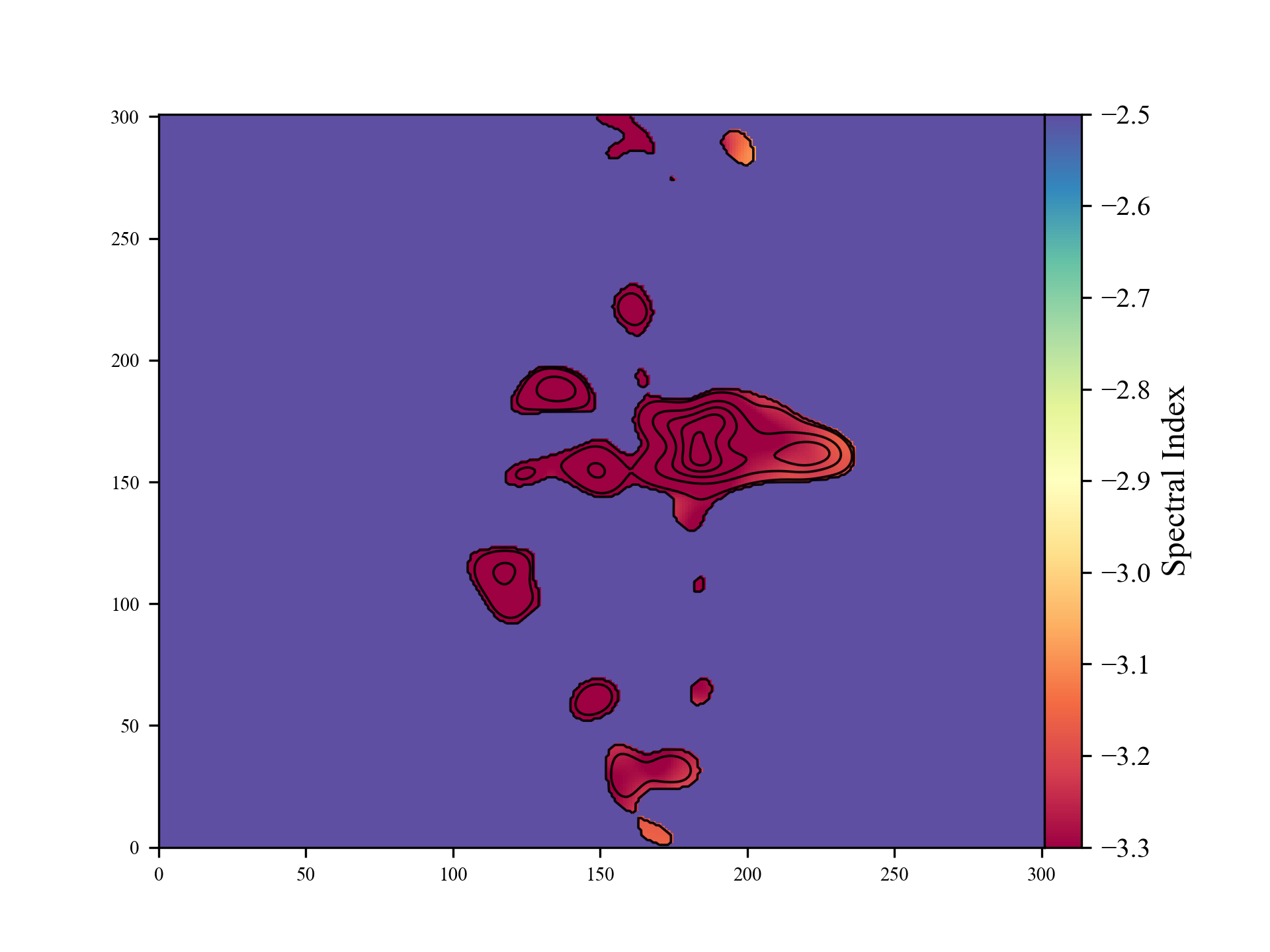}
    \includegraphics[width=1\linewidth,clip,trim={1.1cm 0.8cm 0.8cm 1cm}]{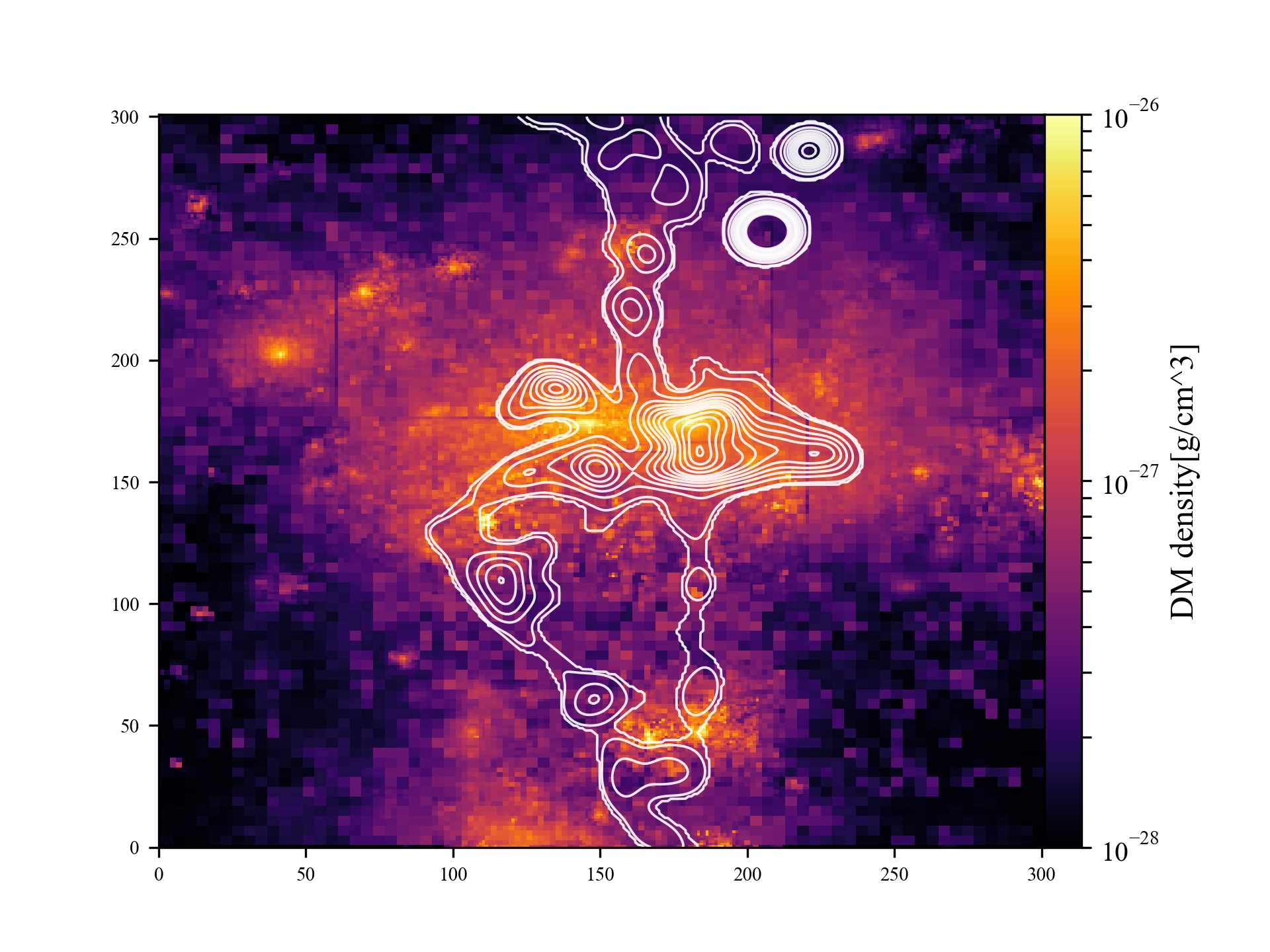}
    \caption{\textit{Top:} Simulated radio emission contours at 120 MHz for our baseline  model for electrons. Contours start at \SI{7}{\milli \jansky \per \beam} and increase by factors of $\sqrt{2}$. The emission has been convolved to the stated resolution of the MWA \SI{118.5}{\mega \hertz} image, \SI{123 x 87}{\arcsecond}, and the spatial scale in the image is \SI{5}{\kilo \parsec \per {pixel}}. \textit{Middle:} Same as in the top panel, but without the inclusion of `cluster weather'. \textit{Bottom:} The dark matter density, where concentrations indicate the presence of galaxies and show the approximate alignment of radio peaks with their original host galaxies.}
    \label{fig:franco}
\end{figure}

To explore a possible formation scenario for the USS Jellyfish, we turn to a recent suite of magnetohydrodynamical simulations by \citet{Vazza2021}. In one of these scenarios, we have identified a polyphoenix akin to the USS Jellyfish as a transient feature in the evolving ICM. In these simulations, light jets (i.e.,\@ filled by hot gas and magnetic fields) were released by AGN particles in the simulation within a forming galaxy cluster with a total mass of $M_{\rm 100} \approx$~\SI{1.5E14}{\solarmass}. For the runs used in this comparison, the AGN jets were initiated at the same epoch for the four most massive AGNs in the simulation, all located within $\sim \rm Mpc^3$ (comoving), and each releasing $\sim$\SI{E57}{\erg} of feedback energy into the surrounding medium in the form of kinetic, thermal, and magnetic energy. The model for the spectral energy evolution of relativistic electrons is similar to the one presented in \citet{Vazza2021}, and we describe it briefly here. The spatial propagation of relativistic electrons injected by AGNs is followed in post-processing using a Lagrangian advection algorithm, as detailed in \citet[][]{wittor20}. The spectral energy evolution is modelled by numerically integrating Fokker-Planck equations of an initial power-law distribution of electron momenta, $N(\gamma) \propto \gamma^{-2}$, in the range of $\gamma_{\rm min}=50$ and $\gamma_{\rm max}=10^5$ (where $\gamma$ is the Lorentz factor of electrons). We consider continuous energy transfer via synchrotron emission, inverse Compton scattering, Coulomb collisions, adiabatic compression/expansion, and injection and reacceleration of electrons from shock induced DSA \citep[e.g.][]{ka12}. The AGNs produced by such a model are compatible with FR II morphologies and start with a radio luminosity of $\sim$\SIrange[range-phrase=--,range-units=single]{E40}{E41}{\erg \per \second \per \hertz} at a \SI{120}{\mega \hertz}. The lobes of these AGNs spatially evolve by diffusing into the ICM and rapidly fade in the radio band due to adiabatic losses and radiative cooling, except when perturbations in the ICM reaccelerate fossil electron spectra via adiabatic compression and, most importantly, DSA processes. Turbulent reacceleration processes have not been included.

By visually inspecting the radio emission maps of all snapshots in the simulation, we identified a short ($\sim$\SIrange[range-phrase=--,range-units=single]{100}{250}{\mega \yr}) evolutionary stage, some \SI{2.2}{\giga \yr} after the original AGN outburst, during which the aging lobes released by three of the four AGNs produced a radio morphology that loosely resembles the USS Jellyfish in extent, total power, and radio spectral slope.\footnote{Movies of the evolution of radio emission in this simulations are visible at \url{https://vimeo.com/491986204} and \url{https://vimeo.com/491983312}.} In the epochs prior, the AGN lobes have rapidly dimmed; however, a confluence of factors causes a transient re-brightening coinciding with the mixing of three of the four ancient lobes at the same time as the passing of several weak shocks ($\mathcal{M} < 2$) triggered both by cluster growth as well as interactions of the AGN lobes themselves. The top panel of \autoref{fig:franco} shows the contours of radio emission at \SI{120}{\mega \hertz} from this snapshot, if the source were placed at the same distance as Abell 2877, with contours marking $\geq$~\SI{7}{\milli \jansky \per \beam} regions, whilst the colours give the spectral index in the \SIrange[range-phrase=--,range-units=single]{87}{150}{\mega \hertz} frequency range. At least qualitatively, the detectable emission in the simulated emission map resembles the complex shape of the real USS Jellyfish, including its elongated threads and radio substructures in the `head'. The spectral index map has a rather uniform distribution with $\alpha \sim -3$, with some patches of steeper emission in the threads. This model takes into account the role of weak shock (re)acceleration in the ICM, driven by matter accretion events that took place in the host cluster after the injection of jets. The middle panel shows the same snapshot but without `cluster weather'---that is, excluding all ICM interactions such as weak shocks induced by gas motions associated with cluster growth---and points to the necessity in this particular case for these external mechanisms to draw out the tentacles and filamentary structures. The bottom panel shows the dark matter density, which as a proxy for the location of galaxies in the simulation, shows that three of the peaks in the radio emission still closely align with their host galaxies, even after significant evolution of the system.

This simulation tentatively suggests that multiple aging and interacting AGN plasmas, alongside weak shock-induced adiabatic compression and DSA, are a plausible and sufficient explanation for diffuse and very steep cluster emission like the USS Jellyfish, and that alternative mechanisms, such as those developed to explain the GReET, are not necessarily required here. The simulation also shows the potential rarity of the scenario due both to its short-lived, transient nature in a lengthy \SI{2.2}{\giga \yr} evolution and the requirement for an otherwise radio-quiet cluster across the remainder of that evolution.

\section{Conclusion}

We have reported on the discovery of the steepest spectrum synchrotron source to date, the `USS Jellyfish', which lies in the northwest of the cluster Abell 2877. We have argued that the source is a polyphoenix: that it is composed of multiple aged and mixed AGN populations reaccelerated by a common large-scale event. The currently available X-ray data do not show the presence of cold fronts, sloshing, or other shock systems in the cool-core cluster; however, the quality of these data likely means that weak shocks are undetectable. We have also presented recent magnetohydrodynamic simulations showing that a combination of weak shocks inducing standard DSA and adiabatic compression and acting upon ancient, interacting AGN plasmas are capable of producing diffuse, USS emission akin to the USS Jellyfish without recourse to other, more exotic (re)acceleration mechanisms. These same simulations show that the USS Jellyfish may also be a short-lived, transient phase in the evolution of the system.

Follow-up work on Abell 2877 should make high-sensitivity and high-resolution X-ray observations of Abell 2877 a priority in an effort to discern the presence and nature of any shocks in the northwest and detect telltale signs of cool-core sloshing that may be present. Additionally, both higher-resolution and higher-sensitivity radio observations of the USS Jellyfish may provide us with a better picture of its morphology; however, the incredible steepness of its spectrum makes follow-up observations above \SI{300}{\mega \hertz} unlikely to detect any of the extended emission. We may have to wait for the development of SKA Low for a higher-resolution, low-frequency telescope able to observe so far south.

\acknowledgements The authors thank Benjamin Quici for helpful discussions during the preparation of this manuscript.
This scientific work makes use of the Murchison Radio-astronomy Observatory, operated by CSIRO. We acknowledge the Wajarri Yamatji people as the traditional owners of the observatory site. Support for the operation of the MWA is provided by the Australian Government (NCRIS) under a contract to Curtin University administered by Astronomy Australia Limited. We acknowledge the Pawsey Supercomputing Centre, which is supported by the Western Australian and Australian Governments.
The Australia Telescope Compact Array is part of the Australia Telescope National Facility, which is funded by the Australian Government for operation as a National Facility managed by CSIRO.
F.V. acknowledges financial support from the ERC Starting Grant MAGCOW, No. 714196. The cosmological simulations in Sec.4
 were performed with the ENZO code (\url{http://enzo-project.org}), under projects ``stressicm" and ``hhh44" at the J\"ulich Supercomputing Centre (PI: F. Vazza).
D.W. is funded by the Deutsche Forschungsgemeinschaft (DFG, German Research Foundation) 441694982.

\bibliographystyle{aasjournal}
\bibliography{refs.bib}

\end{document}